\documentclass[apj]{emulateapj}
\def\lap{\lower.5ex\hbox{$\; \buildrel < \over \sim \;$}}
\def\gap{\lower.5ex\hbox{$\; \buildrel > \over \sim \;$}}

\def\kms{km s$^{-1}$}
\def\m8{m$_{8}$}
\def\m45{m$_{4.5}$} 
\usepackage{graphicx}
\usepackage{natbib}

\begin{document}

\shorttitle{Star Formation in the Bullet Cluster}
\shortauthors{Chung et al.}
\title{\footnote{This paper includes data gathered with the 6.5 meter Magellan Telescopes located at Las Campanas Observatory, Chile.}Impacts of a Supersonic Shock Front on Star Formation in the Bullet Cluster}
\author{Sun Mi Chung\altaffilmark{1}, Anthony H. Gonzalez\altaffilmark{1}, Douglas Clowe\altaffilmark{2}, Dennis Zaritsky\altaffilmark{3}, Maxim Markevitch\altaffilmark{4}, Christine Jones\altaffilmark{4}}

\altaffiltext{1}{Department of Astronomy, University of Florida, Gainesville, FL 32611-2055; schung@astro.ufl.edu}
\altaffiltext{2}{Department of Physics and Astronomy, Ohio University, 251B Clippinger Lab, Athens, OH 45701}
\altaffiltext{3}{Steward Observatory, University of Arizona, 933 North Cherry Avenue, Tucson, AZ 85721}
\altaffiltext{4}{Harvard-Smithsonian Center for Astrophysics, 60 Garden Street, Cambridge, MA 02138}

\begin{abstract}
We use the Bullet Cluster (1E0657-56) to investigate the extent to which star
formation in cluster galaxies is influenced by ram pressure from supersonic
gas (Mach 3) during a cluster merger.  While the effects of ram pressure have
been studied for individual galaxies infalling into galaxy clusters, this
system provides a unique opportunity to investigate the impact of dramatic
merger events on the cluster galaxy population.  In this analysis we use {\it
Spitzer} IRAC data to study star formation. At the redshift of the cluster the 6.2 $\mu$m PAH feature is redshifted into the 8 $\mu$m band, enabling use of the m$_{4.5}$-m$_{8}$ color as a proxy for specific star formation rate.  We find that the color distribution on the two sides of the shock differ by less than 2$\sigma$, and conclude that ram pressure from the shock front has no dramatic, immediate impact on the star formation of cluster galaxies in the Bullet Cluster.
\end{abstract}


\section{Introduction}

It has long been observed that there is a correlation between galaxy
morphology and local density \citep{dressler1980}, known as the
density-morphology relation.  Although there are various physical mechanisms
that can transform star-forming late-type galaxies into quiescent early-type
galaxies, it is unknown whether this transformation occurs mostly in the
galaxy group or galaxy cluster environment.

Preprocessing of galaxies in the group environment has been argued by some
\citep{zabludoff1998,kodama2001} to be a principal mechanism driving the
predominance of the passive galaxy populations observed in galaxy clusters.
In galaxy groups, the velocity dispersions are sufficiently low that processes
such as galaxy-galaxy interactions can quench star formation and
morphologically alter galaxies.  In addition to galaxy-galaxy interactions,
\citet{kawata2008} show that ``strangulation'' is an important mechanism that
leads to quenched star formation in the group environment.  Using a
cosmological chemodynamical simulation, \citet{kawata2008} show that most of
the hot gas in a disk galaxy is stripped away, thus cutting off a source of
new cold gas (some of the hot gas cools to form cold gas), which is necessary
to maintain star formation.

Others have argued that cluster-specific processes are responsible for the
high early-type fraction. The simulations of \citet{berrier2008} argue that
the transformation of late-type to early-type galaxies is mostly attributed to
processes internal to the galaxy cluster environment, rather than the group
environment.  \citet{berrier2008} find that most of the galaxies in a galaxy
cluster are accreted directly from the field, rather than from infalling
galaxy groups. In this case, mechanisms such as galaxy harassment
\citep{moore1996} and ram pressure stripping \citep{gunn1972} that are most
efficient in massive clusters must drive this transformation.

Simulations have shown that ram pressure stripping can remove some or all of
the interstellar medium (ISM) from late-type galaxies as they enter the
cluster potential \citep{roediger2005,mayer2006,bruggen2008}.
Studies of galaxies in clusters with observed HI deficiencies or truncated
disks support the idea that ram pressure stripping can transform a gas-rich
spiral galaxy into an anemic spiral galaxy
\cite[e.g.][]{vollmer2001,vollmer2008}, which may eventually evolve into a
lenticular \citep[e.g.][but see \citet{boselli2006}]{bekki2002}. 
Ram pressure has also been shown to trigger bursts of 
star formation in cluster galaxies \citep{gavazzi2001,gavazzi2003}, and
recent simulations \citep{kronberger2008} predict enhanced star 
formation rates in galaxies undergoing a ram pressure event.

While previous studies have examined the impact of ram
pressure on individual galaxies falling into the cluster potential \citep{cortese2007,chung2007} the Bullet Cluster provides a unique opportunity to examine the effect of ram pressure induced by cluster mergers. Although galaxies in any cluster environment undergo ram pressure as they travel through the ICM, galaxies in a major merger event such as those we observe in the Bullet Cluster will experience a dramatic enhancement in the pressure since $P\propto V^2$. It is thus possible that
this brief transient phase may have a significant impact on the properties of
the cluster galaxy population. The Bullet Cluster is an ideal site in which to
quantify the importance of such merger-induced ram pressure.

In this paper we conduct an initial exploration of the impact of the shock
front upon polycyclic aromatic hydrocarbon (PAH) emission, and hence star
formation, in cluster galaxies.  We analyze Spitzer data taken with the
Infrared Array Camera \citep[IRAC;][]{fazio2004}, in conjunction with optical
spectra observed with the Inamori Magellan Areal Camera and Spectrograph
\citep[IMACS;][]{bigelow1998}.

\section{Geometry and Cluster Properties} 
The Bullet Cluster, at a redshift of $z=0.296$, is a head-on collision between
two clusters of unequal mass, which we call the main cluster and subcluster.
\citet{markevitch2002} [M02] observed a sharply defined bow shock front,
driven by the subcluster and propagating in the gas of the main cluster with a
velocity of 4740$^{+710}_{-550}$ \kms~(Mach number $M=3.0^{+0.45}_{-0.35}$)
(Markevitch et al. 2008, in preparation). Hydrodynamic simulations showed that
the subcluster itself has a lower velocity relative to the main cluster,
$\sim2700$ \kms\ (Springel \& Farrar 2007; see also Milosavljevi{\'c} et
al. 2007).  While the subcluster and the shock front might be expected to move
together, the gravity of the subcluster causes the main cluster's gas in front
of the shock to flow onto the subcluster.  As a result, the shock front has a
higher velocity in the reference frame of that gas inflow, which is what is
measured in X-rays.  \citet{markevitch2004} constrain the inclination angle of
the collision, with respect to the plane of the sky, to $i<8^{o}$, based on
the Mach number of the shock and subcluster relative line-of-sight velocity
($\sim$600 \kms) \citep{barrena2002}.  From the velocity and geometry of the
system, it is computed that the subcluster exited the main cluster core
$\sim$0.15 Gyr ago.

The simple geometry of the Bullet Cluster makes it an ideal system in which to
study how galaxy properties such as star formation are affected by ram
pressure.  Because the merger is nearly in the plane of the sky, the velocity
of galaxies with respect to the intracluster medium (ICM) is well constrained,
providing a good quantitative measure of the ram pressure exerted by the gas
on the galaxies.  In addition, the recentness of the collision (0.15 Gyr)
allows us to use sensitive indicators of recent star formation, direct and
indirect measures of the ionizing flux from O/B stars with lifetimes of 0.1
Gyr.

\section{Physical Assumptions}

The analysis in this paper requires several assumptions related to the star
formation rate and the three-space galaxy distributions for the main cluster
and subcluster.

\subsection{Star Formation Rate}
To probe star formation rates, we use the PAH emission at rest wavelength
$\lambda=6.2\micron$, which at the cluster redshift is observed in the 8$\micron$
IRAC band.  Ultra-violet (UV) photons that are produced by young hot stars are known to excite PAH molecules which then emit in the $\lambda=6.2\micron$ and other infrared bands \citep{leger1984,allamandola1985}.  More recent studies have shown that the $6.2\micron$ luminosity is correlated with total infrared luminosity  \citep{peeters2004,brandl2006}, which is a tracer of star formation rate \citep{kennicutt1998}.  However, PAH emission is not a unique signature of star formation, which can also be excited by visible photons \citep{uchida1998,li2002}, as well as by active galactic nuclei (AGN) \citep{freudling2003}.  
It has also been shown that PAH emission is dependent on metallicity 
\citep{engelbracht2005,smith2007} and thus not a robust star formation rate indicator for galaxy samples that span a large range in metallicity \citep{boselli2004}.

For the Bullet Cluster, both effects are expected to be second order. 
Our data are negligibly affected by AGN contamination (see \S5), and our 
sample is dominated by massive galaxies, for which the metallicity 
variations are not dramatic. Moreover, the metallicity distribution 
should be the same for galaxies on both sides of the shock. The PAH flux 
should nevertheless be considered a rough proxy for star formation rate 
rather than a precision tracer.

To normalize the 8$\micron$ flux, we use the 4.5$\micron$ flux as a proxy for mass.  The 4.5$\micron$ flux originates
mostly from the old red giant stars, and probes the stellar mass of a galaxy
at the cluster redshift.  By using the ratio of 8$\micron$ to 4.5$\micron$
flux or m$_{4.5}$-m$_{8}$ color, we can roughly trace star formation rate per
stellar mass.  Specifically, $8\mu$m emission in excess of that expected from
the Rayleigh-Jeans tail of the cold stellar component is assumed to arise from
star-formation induced PAH emission.  We estimate the level of impact that the
M$\sim$3 shock has had on star formation activity in individual galaxies by
looking at the m$_{4.5}$-m$_{8}$ color as a function of distance from the
shock front.

\subsection{Galaxy Distribution and Projection}
\begin{figure}[h]
\epsscale{1}
\plotone{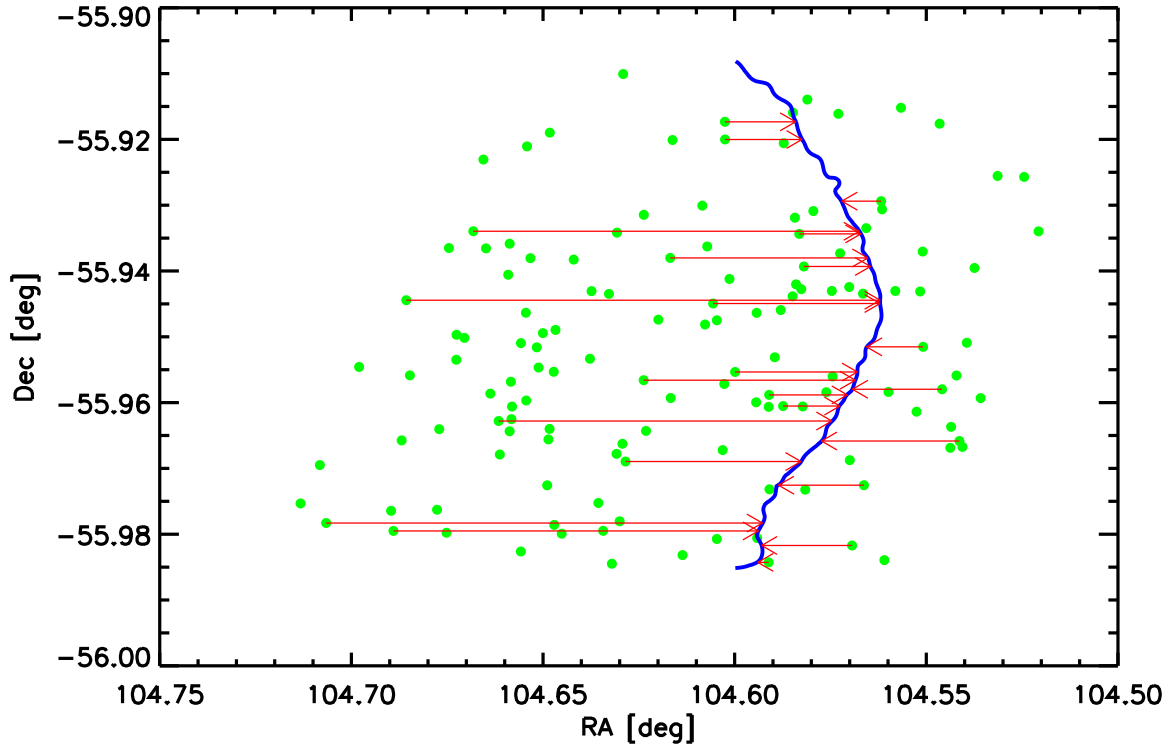}
\caption{Bullet Cluster candidates (filled circles) shown with the shock front (a corresponding contour of constant X-ray surface brightness).  Arrows indicate the distance between the object and the shock front for sources with IRAC color m$_{4.5}$-m$_{8} > 0.5$.}
\label{fig:fig1}
\end{figure}

We proceed with our analysis under the simplifying assumption that galaxies
that lie ahead of the shock front are main cluster ``pre-shock'' galaxies, and
those behind the shock front are main cluster ``post-shock'' galaxies
(Figure~\ref{fig:fig1}).  The post-shock galaxies have already experienced the
ram pressure exerted by the gas behind the shock, whereas the pre-shock
galaxies have not yet been affected.  The true physical situation is clearly
more complex, with contamination from the subcluster galaxies as well as
background/foreground sources not associated with the Bullet Cluster.
Although there is some contamination from the subcluster galaxies, we assume
that the galaxy populations on both sides of the shock front are dominated by
main cluster galaxies based on estimates of the total stellar mass.  The
stellar mass of the main cluster is approximately an order of magnitude larger
than the stellar mass of the subcluster.

We also note that any effect of contamination from the subcluster galaxies
will be strongest at the shock front, which bisects the subcluster brightest
group galaxy.  Subcluster galaxies just ahead of the shock front experience a
strong ram pressure as they collide with the main cluster ICM at $\sim$4700
\kms, whereas subcluster galaxies just behind the shock are somewhat shielded
by the shock front and feel a much weaker ram pressure with a slower gas
inflow velocity of $\sim$1600 \kms.  Therefore any effect of ram pressure on
star formation rate has the opposite sign for subcluster galaxies near the
shock front, compared to the main cluster galaxies.

We assume that there is a sharp division between post-shock and pre-shock galaxies, where all galaxies ``behind'' the shock front (leftward of the curve in Figure \ref{fig:fig1}) 
have experienced the effects of the shock.  However, because of the projection of galaxies and curvature of the shock front surface along the line of sight, there is a gradual transition from post-shock to pre-shock galaxies.
This projection effect softens any step-like increase/decrease in color (or star formation rate) across the shock boundary. 
By generating galaxies in a random distribution according to the Navarro, Frenk, and White (NFW) \citep{NFW1997} profile, and assuming the shock front is an axisymmetric cone projected in the plane of the sky and spanning $\sim$550 kpc from the axis of symmetry, we estimate that the transition region during which the observed galaxies go from being fully post-shock to fully pre-shock galaxies is $\sim$0.5 Mpc.

\section{Observations}
\subsection{IRAC}

Observations of 1E0657-56 were taken on December 17, 2004, with Spitzer/IRAC.
Data were collected from all four IRAC channels - 3.6$\micron$, 4.5$\micron$,
5.8$\micron$, and 8$\micron$, in full array readout mode.  A medium scale,
cycling dither pattern was used, with 20 pointings.  At 2 frames per pointing
and 100 s exposure time per frame, the total integration time per IRAC band
was 4000 s.  The native IRAC pixel scale is 1.22 arcsec, but the final reduced
images were set to have a pixel scale of 0.86 arcsec pixel$^{-1}$, with an
effective field of view (FOV) of 3.7 $\times$ 3.7 arcmin covered by the four
channels.  The frames were processed using the Spitzer Science Center (SSC)
IRAC Pipeline, and mosaics created from the basic calibrated data (BCD) frames
using \emph{MOPEX}.

\subsection{IMACS}
We analyzed optical spectroscopic data taken with Magellan/IMACS in 2005 and
2006.  In 2005, 430 objects were observed with four masks, with exposure times
ranging from 4.8 to 5.4 ks.  During this run, red sequence galaxies were given
highest priority as targets.  The following year, two additional masks were
obtained, with exposure times of 10.8 and 12.0 ks.  Target priority was given
to objects with relatively red m$_{4.5}$-m$_{8}$ colors.  The wavelength range
of all spectra are 4000 \AA$-$9000 \AA, with a dispersion of 2 \AA\ per
pixel. We note that spectroscopic selection criteria for galaxies on both
sides of the shock front are identical.

\section{Data Analysis}
\subsection{IRAC Photometry}

Source detection and photometry of the IRAC images are performed with Source
Extractor (Bertin \& Arnouts 1996).  Sources are identified in the 4.5$\mu$m
image with a 3.5$\sigma$ detection threshold.  Magnitudes are measured within a
5$\arcsec$ aperture.  We choose to apply no aperture corrections to the
photometry.  In this analysis we are primarily interested in colors rather
than total magnitudes, which for m$_{8}$-m$_{5.8}$ and m$_{4.5}$-m$_{3.6}$
would change by only $\sim$0.05 magnitudes with the corrections applied.
Since the corrections themselves are uncertain by up to 10\%
(http://ssc.spitzer.caltech.edu/irac/calib/extcal), excluding systematics, we
use raw aperture magnitudes for all analyses.  In addition, aperture
corrections are expected to be minor when examining colors rather than
absolute magnitudes.

\subsection{IRAC Sample Selection}

The curvature of the shock contour indicates that the direction of propagation
lies within $\sim$5 degrees of the east-west axis (M02). The shock front was
defined as the X-ray brightness contour whose most westward point lies at
104.6 degrees in right ascension (Figure~\ref{fig:fig1}), with a maximum and
minimum declination of -55.908 and -55.985 degrees.  Objects located above or
below (north or south) the shock front were not included in our analyses,
since these sources are not in the direct path of the shock front and thus
should be minimally affected by the propagating shock. We limit our analyses
to the region of spatial overlap between the four IRAC bands, detecting a
total of 758 sources within this ``analysis region''.

We eliminate unsaturated stars from the IRAC catalog using Magellan optical photometry \citep{clowe2006}.  In the inner region of the R-band image
where the point spread function (PSF) is small ($\sim$0.45 arcsec FWHM),
objects with a half-light radius of 2.6 to 3.2 pixels (0.111 arcsec/pixel) and
an R-band aperture magnitude of 18 to 24.1 (with a 7.8 pixel aperture), are
flagged as stars.  Due to a gradient of the PSF in the image, half-light
radius and aperture magnitude cuts are scaled accordingly to provide a catalog
of unsaturated stars.  Saturated stars are identified manually by inspecting
the Magellan R-band image to look for saturated objects.

To minimize contamination from bright foreground galaxies, we impose a
magnitude limit of m$_{4.5} > 14.20$ -- the magnitude of the brightest cluster
galaxy (BCG) in the main cluster.  We also apply a magnitude cut of m$_{4.5} <
17$, which is sufficiently faint to detect most of the 8$\micron$ sources.
While going deeper to m$_{4.5} < 18$ increases the number of cluster
candidates, we opted for a more conservative m$_{4.5} < 17$ magnitude limit so
as to minimize the background contamination.  Most importantly, the analyses
described in this paper produce the same result within error bars, whether we
apply a magnitude limit of m$_{4.5} < 17$ or m$_{4.5} < 18$.

Of the 758 sources, 18 are brighter than the main cluster BCG, 176 are stars,
leaving us with 564 sources.  Eliminating sources fainter than m$_{4.5}=17.0$,
we are left with 169 sources in the direct path of the shock front and 31 that
are above/below the shock front. The distance from likely star-forming
galaxies to the shock front is illustrated by arrows in Figure~\ref{fig:fig1}.

\subsection{IMACS Spectroscopy}
To reduce contamination from background and foreground sources, we reduce a
total of 458 IMACS spectra with IRAF and IDL, then cross-correlate them with
template spectra, using the IRAF task \emph{xcsao} to obtain redshifts.  We
use the same templates as \citet{tran2005}, each spectrum representing a
different type of galaxy -- giant elliptical, E+A, Sb spiral, and an emission
line galaxy (ELG).  Of the 458 spectra, we obtain reliable redshifts for 326
sources.  We augment our spectroscopic data set with 133 redshifts
\citep[][priv. comm]{barrena2002}, which more densely sample the cluster core,
yielding a total of 459 sources with known redshifts.

We next associate the spectroscopic redshifts with IRAC sources for the 114
spectra that lie within the IRAC field-of-view.  Of these 114 sources, 67 are
cluster members with redshifts that lie within $\pm 2000$ \kms~of the mean
cluster redshift ($z=0.296$), and 47 are interlopers.  Sixty-three out of the
67 cluster members, and 42 out of the 47 interlopers, meet the m$_{4.5} < 17$
criteria.  This leaves us with 134 ``cluster candidates'', which include all
IRAC sources that meet our magnitude criteria, are located within the direct
path of the shock front, and exclude all known interlopers.

\section{Results and Discussion}

\subsection{Impact of the Shock} 

We use the m$_{4.5}$-m$_{8}$ color to probe relative star formation rates per
unit stellar mass.  At the redshift of the cluster ($z=0.296$), the prominent
PAH emission band at rest wavelength 6.2$\mu$m falls into the 8.0$\mu$m IRAC
band.  PAH emission, which is excited primarily by UV photons, most often
originates from photodissociation regions of star formation sites
\cite[e.g.][]{Howell2007}.  Although PAH emission is sometimes also associated
with AGN \citep{freudling2003}, only 2 out of 134 cluster candidates in the
direct path of the shock front have IRAC colors within the AGN wedge
\citep{lacy2004,stern2005}.  Both AGN candidates have m$_{4.5}$-m$_{8} < 0.6$,
and thus do not contribute spurious star formation signatures.

\begin{figure}
\epsscale{1}
\plotone{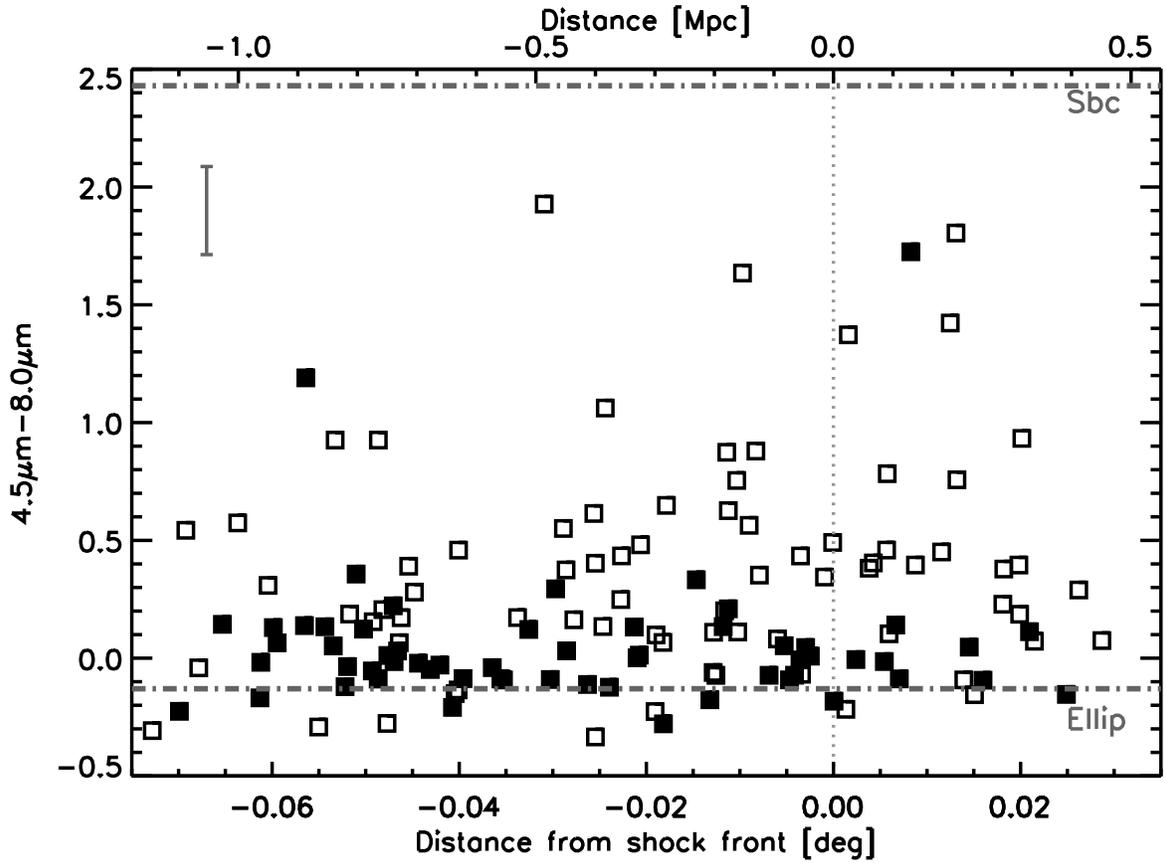}
\caption{The m$_{4.5}$-m$_{8}$ color as a function of distance from the shock front for 134 cluster candidates. The lower and upper axes indicate projected distance in degrees and Mpc, respectively.  Spectroscopically confirmed cluster members are filled in boxes.  The vertical dotted line shows the location of the shock front.  Sources to the left of this line are considered post-shock galaxies, and those to the right are considered pre-shock galaxies.  The horizontal dotted-dashed lines at m$_{4.5}$-m$_{8}$=-0.13 and m$_{4.5}$-m$_{8}$=2.43 represent the expected color of an Elliptical and Sbc galaxy at z=0.3, respectively \citep{assef2007}. The error bar in the upper left corner shows the mean color error of the 134 candidates.}\label{fig:fig2}
\end{figure}

Figure~\ref{fig:fig2} shows color as a function of distance from the shock
front for 134 cluster candidates.  The dotted horizontal lines show the
expected color of an elliptical and Sbc type galaxy at z=0.3
\citep{assef2007}.  These predicted colors come from a set of low-resolution
spectral templates, derived by \citet{assef2007}, using optical and
near-infrared photometry of over 16,000 galaxies in the NOAO Deep Wide-Field
Survey Bo\"otes region, and redshifts from the AGN and Galaxy Evolution Survey
(AGES).
For elliptical galaxies the 8$\mu$m emission comes from the Rayleigh-Jeans
tail of the stellar component, for which the m$_{4.5}$-m$_{8}\sim-0.1$ at z=0.3 
\citep{assef2007}. Colors redder than the elliptical locus are indicative of 
PAH emission in the 8$\mu$m band.

A majority of the IRAC sources have colors consistent with red sequence
galaxies in the cluster -- only 17\% of the cluster candidates have a color
m$_{4.5}$-m$_{8}>0.5$, and 6\% of the sources have m$_{4.5}$-m$_{8}>1$.
Although there is a selection effect in favor of red sequence galaxies with
the spectroscopic data, Figure~\ref{fig:fig2} uses the entire IRAC sample, for
which there is no bias towards quiescent ellipticals.   

\begin{figure}
\epsscale{1}
\plotone{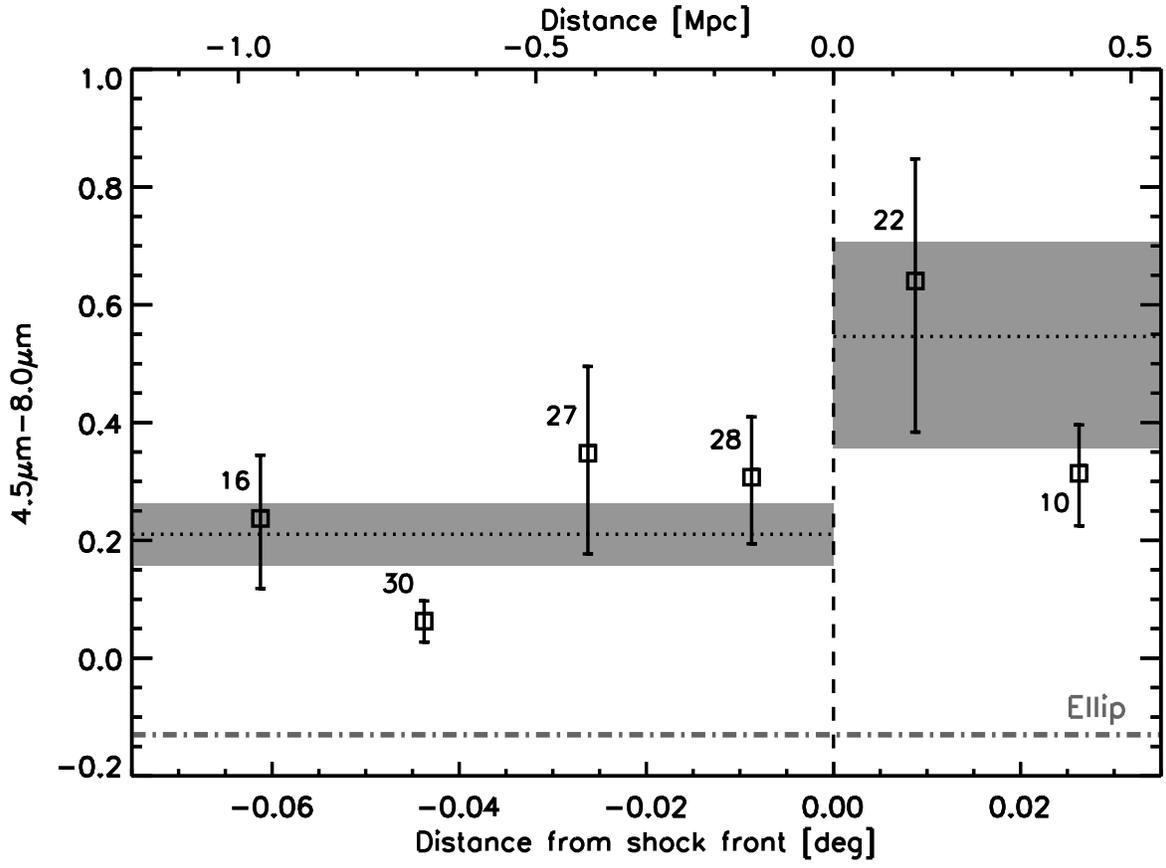}
\caption{The m$_{4.5}$-m$_{8}$ color as a function of distance from shock front for cluster candidates in bins of 275 kpc, with the number of sources per bin denoted.  Individual error bars are derived from bootstrapping method.  Grey shaded regions show the integrated color for all objects behind and ahead of the shock front, $\pm$1$\sigma$ derived from bootstrapping.}\label{fig:fig3}
\end{figure}

The spectroscopic data
is useful for eliminating interlopers, but the sample of confirmed cluster
members (solid points in Figure \ref{fig:fig2} \& Figure \ref{fig:fig3}) 
is of limited value in the
current analysis because the spectroscopic program preferentially targeted red
sequence galaxies.  Indeed, only 2 of the 23 cluster candidates with
m$_{4.5}$-m$_{8}>0.5$, presently have spectra confirming their membership.

Instead, we focus in the current analysis on the full sample of 134 candidate
members and assess whether there is evidence for a correlation between the
luminosity-weighted color of cluster members and their distance from the shock
front.  For each galaxy we compute the distance from the shock (Figure
\ref{fig:fig1}), and then compute the integrated color of all galaxies in 275
kpc wide bins (Figure \ref{fig:fig3}).  For each bin we use the bootstrap
technique with 10,000 realizations to calculate the uncertainties.  The dotted
lines correspond to the total integrated color of all objects on each side of
the shock front, with the width of the shaded grey region corresponding to the
1$\sigma$ confidence interval calculated via the bootstrap method. 

For the individual data points on Figure~\ref{fig:fig3}, the size of the
bootstrap error bars indicates that the mean color is strongly sensitive to
the small subset of strongly star-forming galaxies. For example, in the bin
directly ahead of the shock front, there are 2 (out of 22) galaxies that are
responsible for the relatively high integrated color. One of these two
galaxies is a spectroscopically confirmed, face-on spiral; the other is a
large disk galaxy whose optical color is bluer than the red sequence.  

The horizontal dotted-dashed line in Figure~\ref{fig:fig3} shows the predicted color of an elliptical galaxy at $z=0.3$ \citep{assef2007}.  The data shown in Figure~\ref{fig:fig3} are meant to show an excess of color in the post-shock and pre-shock galaxies, in comparison to what one would expect from a purely quiescent elliptical galaxy.   
Overall, Figure~\ref{fig:fig3} shows that there is no drastic change in color across the shock boundary.  The integrated color of the post-shock versus pre-shock galaxies is consistent to within 2$\sigma$.

The above technique is a rather blunt means of assessing the impact of the
shock front, given that it includes the red sequence galaxies, where we expect
minimal change, as well as the star-forming galaxies.  An alternate approach
is to use the Kolmogorov-Smirnoff (KS) test to understand how statistically
different the populations are on either side of the shock.
We first compare the full distributions on the two sides of the shock front.  
Using all 134 candidate members, a KS-test indicates that the colors on the two sides of the shock differ at only the 1.3$\sigma$ level; consequently, the data do not exclude the hypothesis that the galaxies are all drawn from the same population.
We note that the full sample includes elliptical galaxies, whose m$_{4.5}$-m$_{8}$ color simply reflects the cold stellar component of the Rayleigh-Jeans tail rather than star formation.  All galaxies are included in our initial KS-test because we are looking for changes in the global color distribution of galaxy colors across the shock front.  

To avoid the gas-poor elliptical galaxies whose m$_{4.5}$-m$_{8}$ color is not a good proxy for specific star formation rate, we can exclusively use galaxies whose m$_{4.5}$-m$_{8}$ color is redder than some value.  Comparing the color distribution across the shock front for galaxies with m$_{4.5}$-m$_{8} > 0.2$, m$_{4.5}$-m$_{8} > 0.5$, and m$_{4.5}$-m$_{8} > 1$, the KS-test indicates that the colors on the two sides of the shock front differ at the 0.8$\sigma$, 1.5$\sigma$, and 1.2$\sigma$ level, respectively.  Although we avoid the elliptical galaxies by imposing color cuts, we also introduce an uncertainty since objects may move in and out of the sample due to the impact of the shock front, without leaving a known trace on the final color distribution.  However, we emphasize that the KS-test for all of our color cuts shows no evidence for significant change in the color distribution from pre-shock to post-shock galaxies.   

\begin{figure}
\epsscale{1}
\plotone{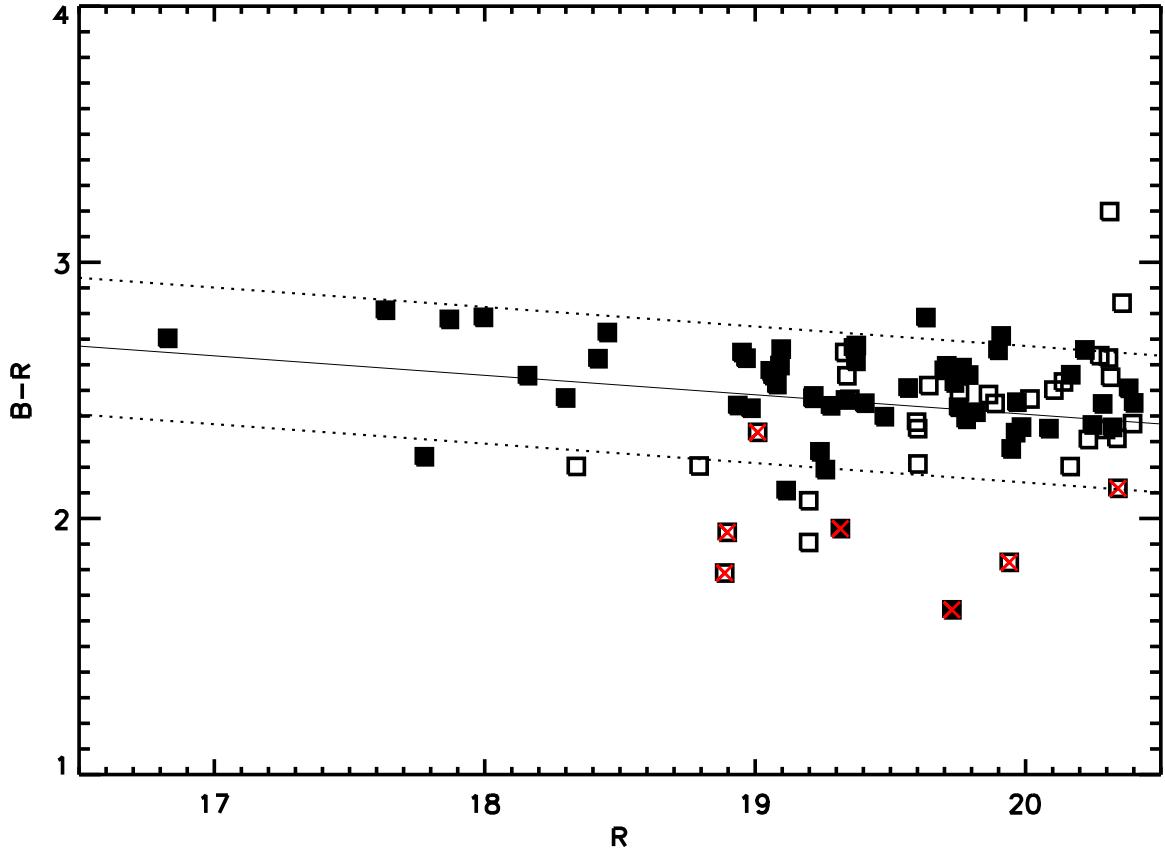}
\caption{The optical color-magnitude diagram shown for 88 IRAC selected cluster candidates.  The solid line is the color-magnitude relation whose slope and zero-point were adopted from \citet{lopez-cruz2004}.  The dotted lines signify 2$\sigma$ from the fit, where the dispersion is obtained by fitting a Gaussian to the residuals.  Objects overplotted with a cross symbol have IRAC color m$_{4.5}$-m$_{8}>0.5$ and are thus likely to be star-forming galaxies.}\label{fig:optical}
\end{figure}

For comparison, we also investigate the optical properties of our IRAC 
sample. Figure~\ref{fig:optical} presents the color-magnitude relation (CMR) 
for the 88 objects out of the 134 IRAC cluster candidates with $R<20.5$, using our Magellan data and again excluding spectroscopically confirmed interlopers. 
As indicated by the IRAC colors, Figure~\ref{fig:optical} confirms that most of the IRAC sources are red sequence galaxies.  
However, to determine exactly which objects can be classified as part of the red sequence, we need three parameters of the CMR -- the slope, zero-point, and dispersion.

We adopt a CMR slope of -0.076, derived from 57 X-ray clusters examined by \citet{lopez-cruz2004}, with redshifts ranging from 0.02$\leq$ z $\leq$0.18.  
To obtain the zero-point of the CMR relation, we fit the data with this fixed slope.  This fit yields a zero-point of 3.927, which is $\sim$0.1 mag brighter than the expected zero-point calculated from the zero-point-redshift relation of \citet{lopez-cruz2004}.  Finally, we calculate the dispersion by fitting a Gaussian function to the CMR 
residuals, excluding outliers.  We derive $\sigma=0.136$ mag, indicating 
that our dispersion is indeed dominated by photometric uncertainty.

We classify all objects within 2$\sigma$ of the CMR as red sequence galaxies, indicated by the dotted lines in Figure~\ref{fig:optical}.  Out of the 7 objects with m$_{4.5}$-m$_{8}>0.5$ and $R<20.5$, 5 of them lie more than 2$\sigma$ below the CMR (with one object just barely within 2$\sigma$).  These red IRAC sources consitute 5/12 or 42\% of the objects below the 2$\sigma$ line.  The optical data thus confirm that objects with red m$_{4.5}$-m$_{8}$ colors correspond to optically blue and likely star-forming galaxies.

There are a few caveats in interpreting the results discussed above.  Foremost, there remains contamination by foreground and background sources for the galaxies that lack spectroscopic confirmation.  Foreground interlopers among the star-forming galaxies may depress the observed significance of any color change across the shock front.  It is unlikely however that removing these foregrounds will create a significant difference in the mean m$_{4.5}$-m$_{8}$ color across the shock boundary.  A total of 55\% of IRAC sources with color 
m$_{4.5}$-m$_{8}>0.5$ are known interlopers, while the fraction of known
interlopers is only 22\% for sources with m$_{4.5}$-m$_{8}<0.5$.  Because the
fraction of interlopers is higher for late-type galaxies than early-type
galaxies, eliminating these objects will likely cause the mean color on both
sides of the shock front (Figure~\ref{fig:fig3}) to approach the quiescent
value.

\subsection{Physical Interpretation} 

The IRAC data show no significant change in color across the shock boundary
(Figure~\ref{fig:fig3}), indicating that the gas behind the shock front does
not have a large impact on the current or recent star formation activity in
the individual galaxies.  
One possible explanation is that ram pressure
stripping may deplete existing gas reservoirs in the outskirts of galaxies
without disturbing current/recent star formation sites that may be more
centrally located in the disk.  Following \citet{gunn1972}, a galaxy moving
face-on through an ICM experiences a ram pressure of $P=\rho V^{2}$.  If ram
pressure exceeds the galaxy's internal gravitational pressure, ram pressure
stripping occurs.  The gravitational pressure can be expressed as $P=2\pi
Gx\frac{M_{star}^2}{r_{g}^4}$, where $G$ is the gravitational constant, $x$ is
the interstellar gas to stellar mass ratio, $M_{star}$ is the stellar mass,
and $r_{g}$ is the galaxy radius.  A galaxy with stellar mass
$M_{\star}=10^{10} M_{\odot}$ and $x=0.6$, should experience ram pressure
stripping at a distance of $\sim$4 kpc from its center, assuming that it moves
face-on through an ICM of density $4\times 10^{-3}$ cm$^{-3}$, at a velocity
of $\sim$2700 \kms. This is somewhat beyond the expected disk scale length for
a galaxy with stellar mass $M_{\star}=10^{10} M_{\odot}$ \citep{pizagno2005}.
Therefore, we might not expect to see evidence of disturbed star formation due
to ram pressure, since much of the current star formation occurs within the
disk scale length.  However, the gas density and galaxy velocity (with respect
to the ICM) are highly uncertain at any significant distance from the shock
front, since both quanities change with time and position in the cluster.

If ram pressure stripping destroys gas reservoirs in galaxies that host
significant star formation, we might expect to see an excess of post-starburst
galaxies in $\sim$10 Myr (after a galaxy crosses the shock front), once the O
stars no longer significantly contribute to the stellar population.  A
timescale of 10 Myr corresponds roughly to 75 kpc in Figure~\ref{fig:fig3}.
On such short timescales, it is difficult to detect an excess of any
particular galaxy type because of small number statistics (i.e.--there are not
enough galaxies within a single 75 kpc bin).  Also, post-starburst galaxies are difficult to distinguish from quiescent elliptical galaxies with photometry, since both populations lie near the optical CMR \citep{tran2007}.

However, the lifetime of a post-starburst galaxy is $\sim$1 Gyr
\cite[e.g.][]{quintero2004}, providing us with a longer timescale (or distance
from the shock front) in which we can search for an excess of these galaxy
types.  In future work, we will spectroscopically identify the E+A galaxies in
our sample.  A higher fraction of E+A galaxies behind the shock front versus
ahead of the shock front, coupled with no change in current star formation
rates across the shock boundary, would favor ram pressure stripping of gas
reservoirs, rather than some effect on current star formation.

Although Figure~\ref{fig:fig3} shows that there is no significant change in
color in the pre-shock versus post-shock galaxies, the geometric effects and
contamination from non-cluster sources may dilute the signal.  Future work
includes analysis of IRAC and MIPS 24$\mu$m data observed over a wider area, 
which coupled with spectroscopic redshifts will be less
sensitive to geometric effects and contamination.  However, even with our
current level of contamination, it is clear that the ram pressure exerted from
the supersonic gas does not have a dramatic impact on current star formation
activity in the main cluster galaxies.

Another source of signal dilution is the large fraction of early-type galaxies
in our sample, for which there are two possible explanations.  First, galaxy
preprocessing in the group environment may have already done its work on both
the cluster and subcluster galaxies, explaining the observed lack of late-type
galaxies \citep{zabludoff1998, cortese2006}.  Second, the galaxies we observe
may have been affected by the cluster environment in such a manner that they
have already converted their gas reservoir into stars.  \citet{marcillac2007}
and \citet{bai2007} have investigated star formation of galaxies in merging
clusters at $z\sim$0.83.  They find evidence of triggered star formation in
infalling galaxies and galaxy groups, which coupled with mechanisms such as
ram pressure stripping or galaxy harassment, can act to eventually quench star
formation.  In this scenario, many of the initially gas-rich, star forming
galaxies in the main cluster would have consumed much of their available gas
in an earlier epoch of triggered star formation, explaining the lack of
late-type galaxies in our current sample.  However, the relative lack of
late-type galaxies cannot explain why we do not see a signal across the shock
front, especially considering that even among the late-type population, we do
not detect a significant difference in color in the post-shock versus
pre-shock galaxies.
  
\section{Summary} 
To trace specific star formation rate, we use IRAC color m$_{4.5}$-m$_{8}$.
The 8$\micron$ flux traces star formation via the PAH emission feature at rest
wavelength 6.2$\micron$, which is redshifted into the 8$\micron$ band at the
cluster redshift.  The 4.5$\micron$ flux can be taken as a proxy for stellar
mass.

Using spectroscopic redshifts, we confirmed 63 out of 200 IRAC sources (169 in
direct path of shock front + 31 above/below the shock front) as cluster
members, and found 42 IRAC sources that are interlopers.  Applying a magnitude
cut at m$_{4.5}=17$, we are left with 158 cluster candidates, of which 134 are
in the direct path of the shock front.

We observe no significant trend in m$_{4.5}$-m$_{8}$ as a function of distance
from the shock front. The KS-test reveals that the sample of galaxies behind the shock versus ahead of the shock front only differs in color by less than
1.5$\sigma$, for galaxies with various cuts in m$_{4.5}$-m$_{8}$ color.  This
indicates that even in a dramatic merger event like the Bullet Cluster, the
ram pressure induced from a supersonic collision does not drastically trigger
or quench current/recent star formation.  A possible explanation is that the
ram pressure can remove the gas reservoirs in the outer disk, while preserving
star formation in the central disk of a galaxy.  While geometric effects,
contamination from interlopers, and the lack of late-type galaxies in the
Bullet Cluster could contribute to diluting a signal, they are unlikely to
explain the complete lack of signal we observe.  Even among our small sample
of late-type galaxies, we detect no change in color between the pre-shock and
post-shock galaxies to within $1.5\sigma$.

\acknowledgements The authors would like to thank Rafael Barrena for sharing
his redshift data with us.  We also acknowledge support for this work from
NASA/Spitzer grant 1319141.  MM was supported by NASA contract NAS8-39073 and
Chandra grant GO8-9128X.  Christine Jones acknowledges support from Spitzer
Contract 1265584.

{\it Facilities:} \facility{Spitzer (IRAC)}, \facility{CXO (ACIS-I)}, \facility{Magellan:Baade (IMACS)}


\end{document}